\documentclass[12pt]{article}
\usepackage{amsmath}
\usepackage{amsthm}
\usepackage{graphicx,psfrag,epsf}
\usepackage{enumerate}
\usepackage[comma]{natbib}
\usepackage{verbatim} 
\usepackage{url} 
\usepackage{subfigure}
\usepackage{multirow,multicol}
\usepackage{booktabs}
\usepackage{appendix}

\newcommand{\normal}{{\mathcal{N}}}
\newcommand{\Z}{{\bf X}}

\theoremstyle{plain}
\newtheorem{thm}{Theorem}[section]

\begin{document}

\def\spacingset#1{\renewcommand{\baselinestretch}%
{#1}\small\normalsize} \spacingset{1}


\title{\bf Multiple Imputation Using Gaussian Copulas\thanks{ Accepted for publication at \textit{Sociological Methods \& Research}. 
    Florian M. Hollenbach is an Assistant Professor, Department of Political Science, Texas A\&M University, College Station, TX 77843-4348 (email: fhollenbach@tamu.edu); Iavor Bojinov is a PhD Student, Department of Statistics, Harvard University, Cambridge, MA 02138 (email: bojinov@fas.harvard.edu); Shahryar Minhas is an Assistant Professor, Department of Political Science, Michigan State University, East Lansing, MI, 48824 (email: minhassh@msu.edu); Nils W. Metternich is a Senior Lecturer, Department of Political Science, University College London, London, UK WC1H 9QU (email: n.metternich@ucl.ac.uk); Michael D. Ward is a Professor, Department of Political Science, Duke University, Durham, NC 27708 (email: michael.d.ward@duke.edu); and Alexander Volfovsky is an Assistant Professor, Department of Statistical Sciences, Duke University, Durham, NC 27708 (email: av136@stat.duke.edu). This project was partially supported by the the Office of Naval Research (holding grants to the Lockheed Martin Corporation, Contract N00014-12- C-0066). Nils W. Metternich acknowledges support from the Economic and Social Research Council (ES/L011506/1). The work was completed while Alexander Volfovsky was supported by a NSF MSPRF under DMS-1402235. For helpful insights we thank Philippe Loustaunau, among the first of our colleagues to encourage this effort. Stephen Shellman was a strong critic who deserves our thanks too: his criticisms helped us to improve our approach. John Ahlquist, Matt Blackwell, Andreas Beger, Cassy Dorff, Gary King, and Jacob Montgomery provided helpful comments on previous versions of this paper.}}
  
  \author{Florian M. Hollenbach\thanks{Corresponding author
    }\hspace{.2cm}\\
    Department of Political Science, Texas A\&M University\\
    and \\
    Iavor Bojinov\\
    Department of Statistics, Harvard University\\
    and\\
    Shahryar Minhas\\
    Department of Political Science, Michigan State University\\
    and\\
    Nils W. Metternich\\
    Department of Political Science, University College London\\
    and\\ 
    Michael D. Ward\\
    Department of Political Science, Duke University\\
     and\\ 
    Alexander Volfovsky\\
    Department of Statistical Science, Duke University\\}
    
  \maketitle

\clearpage
\begin{abstract}
Missing observations are pervasive throughout empirical research, especially in the social sciences. Despite multiple approaches to dealing adequately with missing data, many scholars still fail to address this vital issue. In this paper, we present a simple-to-use method for generating multiple imputations using a Gaussian copula. The Gaussian copula for multiple imputation \citep{hoff:2007} allows scholars to attain estimation results that have good coverage and small bias. The use of copulas to model the dependence among variables will enable researchers to construct valid joint distributions of the data, even without knowledge of the actual underlying marginal distributions. Multiple imputations are then generated by drawing observations from the resulting posterior joint distribution and replacing the missing values. Using simulated and observational data from published social science research, we compare imputation via Gaussian copulas with two other widely used imputation methods: \texttt{MICE} and \texttt{Amelia II}. Our results suggest that the Gaussian copula approach has a slightly smaller bias, higher coverage rates, and narrower confidence intervals compared to the other methods. This is especially true when the variables with missing data are not normally distributed. These results, combined with theoretical guarantees and ease-of-use suggest that the approach examined provides an attractive alternative for applied researchers undertaking multiple imputations.
\end{abstract}

\noindent%
\vfill

\newpage
\spacingset{2} 
\section{Introduction}
\label{sec:intro}

Missing data problems are ubiquitous in observational data and common among social science applications. Statistical inference that does not adequately account for the missing data is widely known to lead to biased results, and inflated (or deflated) variance estimates \citep{rubin:1976,king:honaker:etal:2001,white2010bias,molenberghs2014handbook}.  
Even though most statistical software platforms provides methods that adequately handle missing data (the most popular of these is multiple imputations (MI)), they are often ignored by applied researchers.\footnote{Principled approaches to missing data have existed for over three decades. First formalized by \cite{rubin:1976}, the number of readily available statistical softwares to deal with missing data has rapidly grown since the 1990s \citep[e.g.][]{king:honaker:etal:2001,honaker:king:2010,vanbuuren:groothuis-oudshoorn:2011,kropko:etal:2014}. Further, see the special issue on the \textit{State of Multiple Imputation Software} in the \emph{Journal of Statistical Software} in 2011 \citep{yucel:2011}.} 

\begin{figure}[ht]
  \includegraphics[width=1\textwidth]{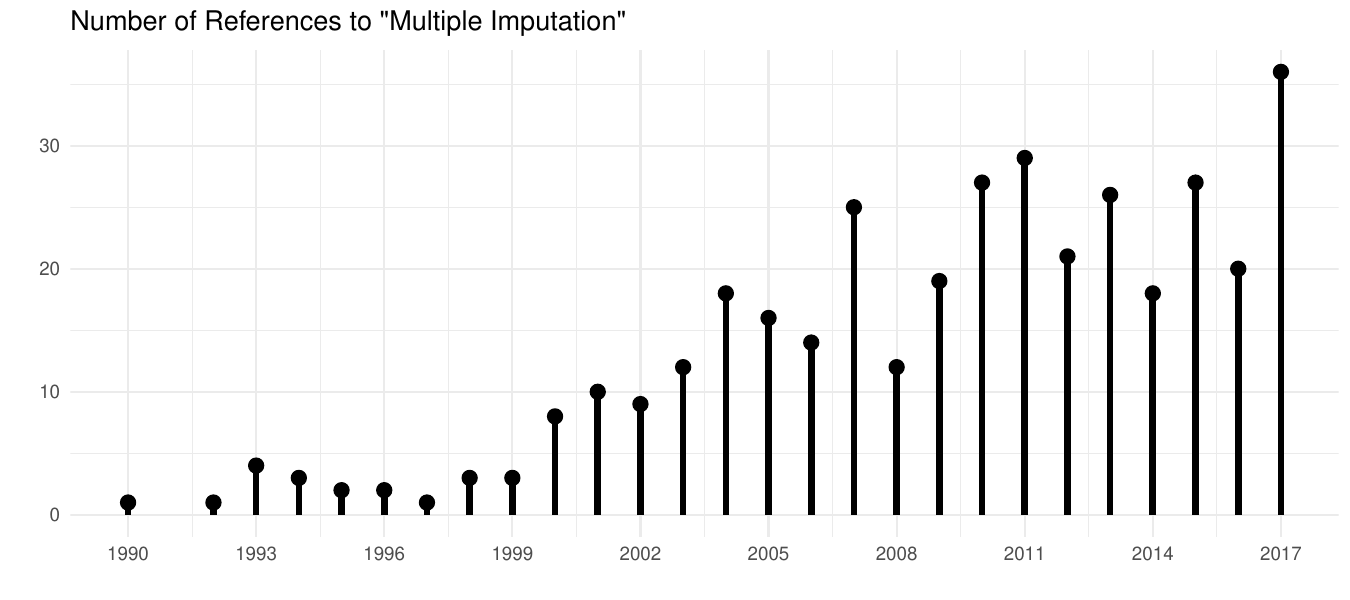}
  \caption{Number of references to ``multiple imputation'' in articles from five top sociology and political science journals since 1990.}
  \label{fig:litCount}
\end{figure}

In Figure~\ref{fig:litCount}, we illustrate the number of articles published in five top sociology and political science journals since 1990 that contain ``multiple imputations'' in the body of the paper.\footnote{The five journals we reviewed from sociology are \textit{Annual Review of Sociology}, \textit{American Sociological Review}, \textit{American Journal of Sociology}, \textit{Sociological Methodology}, and \textit{Sociological Methods \& Research}. In political science we examined the \textit{American Political Science Review}, \textit{American Journal of Political Science}, \textit{Political Analysis}, \textit{British Journal of Political Science}, and the \textit{Journal of Politics}.} Our survey of the literature shows the rapid growth of the use of multiple imputations in the social sciences.  Nevertheless, as missing data is a feature of almost any observational data set, the annual counts of articles mentioning multiple imputations per year still point to significant underutilization of this method in the social sciences. This may be due to a lack of understanding of the benefits (and assumptions) of common imputation methods.
%
%

This article has two aims. First, we introduce applied researchers in the social sciences to a specific copula method for imputation and discuss its advantages over other methods. The method discussed is easy to implement using the \texttt{sbgcop} package \citep{hoff2010package} in $\mathcal{R}$ \citep{r-development-core-team:2004} \footnote{For inexperienced users, our \texttt{gcImp} (\url{https://github.com/bojinov/gcImp}) package provides a simple interface for generating imputations using \texttt{sbgcop}.} 
and has theoretical properties that make it attractive. Second, we conduct a systematic evaluation and comparison of the copula method to two commonly used imputation software packages (\texttt{MICE} \cite{buuren2011mice} and \texttt{AMELIA II} \citep{honaker:etal:2012}) in sociology and political science. 

Copulas are often used for the estimation of dependency between variables and are particularly useful in the generation of imputations as they allow for the construction of valid joint distributions of the data, even if the researcher has little knowledge about the actual joint distribution of the variables. Given the joint distribution of the data, we can generate imputations by sampling from the conditional distribution of the missing data given the observed data. 

We highlight a semi-parametric Gaussian copula approach to missing data imputation. The Gaussian copula is one particular way of constructing a joint distribution from which missing values can be easily drawn. The method was initially developed by \citet{hoff:2007} to estimate empirical models on multivariate data. 

In particular, the Gaussian copula defines the dependence among the distributions of a set of variables which may contain missing values. These variables can include normal, ordinal, and binary variables. Rather than using the distributions themselves, a rank likelihood approximation is used. As a result, the technique does not require the specification of marginal or conditional distributions. This is in stark contrast to other imputation methods using copulas that either require knowledge of the marginals or correlation structure \citep{kaarik2006imputation,kaarik:kaarik:2009, robbins:etal:2013} or target different copula parameters via pseudolikelihood methods \citep{dilascio:etal:2015}. The proposed approach allows applied researchers to undertake imputations of their data without relying on pre-specification or ad-hoc decisions.

The potential use of copulas for multiple imputation applications has not been thoroughly discussed within the social sciences. The copula methods we describe are easy to use and are more likely to provide a good representation of the joint distribution of the data than existing methods. Moreover, provided the Markov Chain Monte Carlo (MCMC) converges, the output from the copula model represents a valid posterior density. Simply put, this means that we have theoretical guarantees about the posterior distribution from which the imputations are generated that other methods can not provide. Based on an extensive simulation exercise, we show that the method presented here is generally at least as accurate as other commonly used methods---it is often better. It also provides better uncertainty estimates for the imputations. Lastly, as is shown in \citet{bojinov2017diagnosing}, the copula method can also be used to test some of the underlying assumptions about the appropriateness of imputations for a given data set.
 
\section{Common Approaches to Multiple Imputation}

The standard techniques employed to deal with missing data require an assumption regarding the missing data pattern; these were first formalized in \cite{rubin:1976}.\footnote{\cite{little:rubin:2002} provide a more up to date treatment and \cite{mealli2015clarifying} an in-depth discussion on the different missing data mechanisms.} To briefly summarize these terms, missing data are missing completely at random (\textbf{MCAR}) when the probability of the observed missing data pattern is unchanged regardless of what values both the observed and missing data take \citep{marini1980maximum}. The missing data are missing at random (\textbf{MAR}) when the probability of observing the missing data pattern is unchanged no matter what values the missing data take. Finally, the missing data are missing not at random (\textbf{MNAR}) when the probability of observing the missing data pattern changes for some values of the missing data. 

These definitions are important both from a theoretical and a practical point of view. The most basic methods, such as listwise deletion, generally lead to biased regression coefficients if the missingness process is not \textbf{MCAR} \citep{graham:2009}. To achieve valid inference under the Bayesian and likelihood paradigms, while ignoring the missing data mechanism, we require the weaker \textbf{MAR} assumption.\footnote{A further assumption of parameter distinctness---the parameter governing the data and the parameter governing the missingness mechanism are \textit{a priori} independent---is required to ensure that valid statistical inference whenever the data are \textbf{MAR} or \textbf{MCAR}. See \cite{little:rubin:2002} for more details on this assumption.}

The most common appropriate approach to dealing with missing data is multiple imputation (MI), which refers to any method that replaces the set of missing values with various plausible values, thus obtaining $m$ completed data sets \citep{rubin1996multiple}.
\citet{rubin:1987} initially suggested creating five imputations, but more recently authors recommended using closer to twenty imputations \citep{vanbuuren:2012}.\footnote{This was based on examining large sample relative efficiency when using a finite number of proper imputations compared to an infinite number, from a Bayesian Gaussian model. In practice, non-normal data combined with non-Bayesian methods can lead to a decrease in the relative efficiency.}  The completed data sets are then separately analyzed using the standard full data techniques and the resulting quantities of interest from each data set are combined to obtain an overall, average estimate as well as its associated variance. 

Before moving to introduce the copula method below, we briefly outline two important methods for generating multiple imputations here. 
\begin{description}
\item[MI with EM] This approach uses iterative expectation maximization (EM) to create complete data sets based on assuming a particular joint distribution. 
A widely used method for imputation in the social sciences is the \texttt{Amelia II} $\mathcal{R}$ package by \citet{honaker:king:2010}  In \texttt{Amelia II}, the joint distribution of the data is modeled as a multivariate normal distribution. \texttt{Amelia II} provides an implementation of the EM approach by the use of bootstrapping to derive solutions quickly. One of the disadvantages of imputation via EM is that for large data sets with significant amounts of missing data, it is computationally intensive. This is a trait of EM algorithms in general, as the rate of convergence is proportional to the amount of missing information in the model. Moreover, it is often unclear to what degree modeling the joint distribution of the data as a multivariate normal distribution is appropriate, especially since the data may include binomial and ordinal variables. 
 
 \item[Conditional Approaches to Multiple Imputation]
An alternative method is to model each variable's imputation via its conditional distribution based on all other variables in the data. One such approach is developed in Multiple Imputation via Chained Equations (MICE) \citep{vanbuuren:2012}, another was developed as the \textit{MI} package in $\mathcal{R}$ \citep{goodrich:etal:2012}. Imputations for fully conditional specification (FCS) methods, such as \texttt{MICE} or \texttt{MI}, are created based on an ``appropriate generalized linear model for each variable's conditional distribution'' \citep[501]{kropko:etal:2014}. This is done for all variables and iterated until the model converges. 

One of the main drawbacks of the FCS is that only under certain conditions do the individual conditional models define a valid joint distribution. This often leads to pathologies in the convergence of the algorithms \citep{li2012imputing,chen2015behaviour}. For example, if $Y|X$ is specified to be an Exponential random variable with rate $X$ and $X|Y$ is specified to be an Exponential random variable with rate $Y$, it is well known that no joint distribution exists and sequentially sampling from these two distributions generates draws that tend to infinity \citep{casella:george:1992}. More strikingly, Example 1 of \citet{li2012imputing} demonstrates that even when all the conditionals are normal, the order in which the variables are updated in MICE can determine whether the chain will converge to a stationary distribution.

One of the advantages of conditional model specification is that it allows each variable to be modeled based on its specific distribution, which is specified by the researcher. However, this also means the imputation model for each variable in the data has to be correctly specified, which can be ``labor-intensive and challenging with even a moderate number of variables'' \citep[41]{murray:2013}. Moreover, coefficients estimates in the conditional models can suffer significantly when the number of missing observations is large, especially for categorical variables \citep{murray:2013}.
\end{description}

\section{A copula approach to missing data imputation}

One of the key issues with conditional approaches to imputation, such as \texttt{MICE}, is that they do not necessarily specify a valid joint distribution (such as the example in the previous section).\footnote{Some theoretical results for \texttt{MICE} are available, but they do not allow too much misspecification in the conditional models. For example, \citet{Liu2013} showed that for valid semicompatible models (\emph{i.e.}, models which are compatible when some of the parameters in the conditional distributions are set to zero, and the joint model obtained from the compatible conditionals contains the correct joint distribution) the combined imputation estimator is consistent. Further, \citet{zhu2015convergence} extend these results to more incompatible models at the expense of the type of missingness patterns allowed (restricting the theoretical results to missingness patterns where each individual is missing at most one variable).}  When a valid joint distribution does not exist, then there are no guarantees that the MI procedure is proper (as defined in \citealp{rubin2004multiple}). A natural approach to overcoming a possibly incompatible conditional specification is by specifying the joint distribution directly. For example, this is done in most EM approaches, such as \texttt{Amelia II}, by simply assuming a multivariate normal distribution. However, while an approximation, most social science data include binary and ordinal variables, and thus cannot have a multivariate normal joint distribution. As a result, this misspecification of the joint distribution is problematic. Moreover, specifying the correct joint distribution becomes increasingly complicated as the number of covariates in the model increase. 

It is therefore valuable to decouple the specification of the marginal distribution of each covariate from the function that describes the joint behavior of all covariates together. One of the main advantages of using copulas for imputation is that they allow us to do exactly that. Sklar's \citeyearpar{sklar:1959} theorem guarantees that every joint distribution can be decomposed in this way:
\begin{thm}[Sklar's Theorem]
	Let $F$ be a $p$-dimensional joint distribution function with marginals $F_1,\dots,F_p$. Then there exists a copula $C$ with uniform marginals such that
	$$F(x_1,\dots,x_p) = C(F_1(x_1),\dots,F_p(x_p))$$
\end{thm}
Sklar's theorem guarantees that the function $C$ is unique if the marginal distributions $F_1,\dots,F_p$ are continuous. If they are discrete, then it is unique on the cross product of the ranges of the $F_j$. 

Much work has been done studying the class of Gaussian copulas where the multivariate dependence is defined by $C$ via the multivariate normal distribution with a correlation matrix $R$ \citep{klaassen1997efficient,pitt2006efficient,chen2006efficient,hoff:2007}. That is, we define the Gaussian copula function as $C(\cdot|R)=\Phi_p(\Phi^{-1}(u_1),\dots,\Phi^{-1}(u_p)|R)$ for $u_1,\dots,u_p\in(0,1)^p$ where $\Phi$ is the univariate normal CDF and $\Phi_p(\cdot|R)$ is the $p$-dimensional CDF with correlation matrix $R$. This means that the joint distribution of the $p$ variables is given by $F(x_1,\dots,x_p) = \Phi_p(\Phi^{-1}(F_1(x_1)),\dots,\Phi^{-1}(F_p(x_p))|R)$. Simply put, the univariate CDFs $F_1,\dots,F_p$ of the individual variables are bound together as a multivariate normal CDF where $R$ determines the correlation between the individual variables on the normal scale.

As previously noted, the specification of marginal distributions is difficult in applied settings and so of particular interest is the setting where the researcher does not need to specify the marginal distributions for $F_1,\dots,F_p$. In fact, one big advantage to the method discussed here is that we consider a semiparametric approach that does not require parameterizing the $p$ marginal distributions.

In this flexible setting, the estimation procedures described below provide consistent and likely asymptotically efficient estimates of the dependence parameters in the Gaussian copula, \emph{i.e.}, $R$ above \citep{murray2013bayesian,hoff2014information}. These dependence parameters directly impact the imputation of the missing data, and thus these theoretical results are extremely appealing. The estimation approach we explore below was developed by \citet{hoff:2007} by extending the ideas of the rank likelihood of \cite{pettitt1982inference} to the copula setting. 

The rank likelihood \citep{pettitt1982inference} is a type of marginal likelihood that bases inference on the ranks of data rather than the full data. In a univariate setting it is defined as follows: consider $z_1,\dots,z_n|\theta\sim p(z|\theta)$ be a sample from some distribution. Instead of observing the actual values $z_1,\dots,z_n$, however, consider only observing the ordering of the data $x_1,\dots,x_n$ (\emph{i.e.} their rank). Then the rank likelihood is given by $$L(\theta;x_1,\dots,x_n) = \int_D p(z_1,\dots,z_n|\theta)dz_1,\dots,dz_n$$ where $D=\{z_{\alpha_1}<\cdots<z_{\alpha_n}\}$ and $\alpha_i=j$ if and only if $z_j$ is the $i$th smallest of $z_1,\dots,z_n$.

\citet{hoff:2007} extends the rank likelhood to the multivariate setting by considering the semiparametric Gaussian copula. Let $z_1,\dots,z_n|R\sim N(0,R)$, with $z_i = (z_{i1}, \dots, z_{ip})$, and let $x_{ij} = F_j^{-1}(\Phi(z_{ij}))$. That is, latent data are drawn from a multivariate normal distribution with correlation $R$ and are transformed to the observed scale via an inverse transformation as in the definition of the Gaussian copula above. One can consider the observed data as the ranks of the unobserved latent $Z$s and define 
$$D=\{Z\in R^{n\times p}:\max\{z_{kj}:x_{kj}<x_{ij}\}<z_{ij}<\min\{z_{kj}:x_{ij}<x_{kj}\}\}.$$
It is easy to see that all $Z\in D$ respect the order of the variables on the observed scale. \citet{hoff:2007} shows that $P(Z\in D|R,F_1,\dots,F_p) = P(Z\in D|R)$ which in turn allows for the decomposition
$$P(X|R,F_1,\dots,F_p) = P(Z\in D|R)P(X|Z\in D,F_1,\dots,F_p).$$ The aforementioned results guarantee that inference about $R$ can proceed simply via $P(Z\in D|R)$. This leverages the ordering of the observed values $x_{1j},\dots,x_{nj}$ of each variable to make inference about the parameter $R$ without estimating the CDFs $F_1,\dots,F_p$.

This means that regardless of the marginal distributions of the individual variables, all we need is their ordering to facilitate the use of the Gaussian copula model to make inferences about the dependence between these variables, \emph{i.e.}, the correlation matrix $R$. A Bayesian approach to estimating $R$ specifies an inverse Wishart prior for a covariance matrix $V$ such that $R$ is its correlation matrix and a normal prior for the latent $z_{ij}$. Updates are performed via a Gibbs sampler since full conditional distributions can be derived by conditioning on the ranks of the data alone.\footnote{Further details of the algorithm for estimation are available in \citet{hoff:2007}.}

Let us paraphrase and summarize the method in less technical terms. Assume we have two vectors $Z_1$ and $Z_2$ which come from a bivariate normal distribution with correlation $R$. We observe $X_i=F_i^{-1}(\Phi(Z_i))$ implying that $X_i$ is distributed according to $F_i$. If the $F_i$ are continuous and known, we can recreate the vectors $Z_1$ and $Z_2$ by using the pseudo-inverse CDFs on the original data ($Z_1=\Phi^{-1}(F_1(X_1))$ and $Z_2=\Phi^{-1}(F_2(X_2))$). We could then generate a good estimate of $R$ using the transformed vectors $Z_1$ and $Z_2$ and maximum likelihood estimation, for example $\Sigma_{i=1}^N \frac{Z_1^{(i)} Z_2^{(i)}}{N}$ would be a natural estimate for the correlation. However, when either vector is not continuous, the simple pseudo-inverse transformation does not allow for correct estimation of the correlation. Now assume $X_2$ is a binary or ordinal variable as in many of our cases but that the marginal is not known. Instead of using a plug in value for $Z_2$ (say, by estimating the marginal $F_2$), we contend that the ranks of our latent continuous $Z_2$ are the same as those of the observed variable $X_2$. The estimation procedure then iterates the following two steps: Using the ranks of $X_1$ and $X_2$ and the current estimate of the multivariate correlation $R$, we can draw values of the latent variables ($Z_1$ and $Z_2$) that preserve the rank ordering of the observed data. The second step uses the sampled underlying latent variables to sample the correlation $R$. These steps are iterated until stationarity is reached. Relying on the ranks and latent scale allows us to not specify the marginal distributions of the individual variables and still arrive at a proper solution to estimating $R$.

When values of $x_{ij}$ are missing at random, imputation can be performed first on the latent $z_{ij}$ scale (since the latent variables are normal, sampling from the conditional distribution of the missing data given the observed data requires a multivariate normal draw) and are then transformed to the observed scale using the empirical cumulative density functions. 
As this is a Bayesian procedure we produce a posterior for the missing data. To make our approach comparable to the standard conditional approaches we only employ a few samples from this posterior and use those as multiply-imputed datasets. However, it is natural to consider posterior predictive distributions of parameters of interest or other posterior summaries on a case-by-case basis. For example, the conditional independence graphs of \citet{hoff:2007} succinctly summarize the relationships among many variables.

\section{Comparing \texttt{Amelia II}, \texttt{sbgcop}, and \texttt{MICE}}\label{sec:comparison}

In this section, we compare the working properties of the copula based imputation with those of \texttt{Amelia II} and \texttt{MICE} packages. We evaluate each method based on an extensive simulation study as well as an empirical example from the social sciences, discussed in the next section.

\subsection{Evaluating Imputations}\label{sec:evaluation}

Multiple imputation procedures are specifically designed to yield valid statistical inference (meaning, asymptotically unbiased with correct standard errors and coverage) for population quantities of interest. Since correct estimation of the coefficients and standard errors is critical for obtaining valid statistical inference, any analysis of MI procedures must focus on studying its frequentist properties. Properties such as empirical coverage, average bias, and average interval length of the estimate of the scientific estimand over repeat samples will be of cardinal interest.

We therefore use the following approach to assess the validity of an MI procedure through simulation:
\begin{enumerate}
    \item Define a full data quantity of interest, $\theta$. In our setting, $\theta$ is a set of regression coefficients. 
    \item Generate a complete data set and apply a pre-specified missing data mechanism to remove some observations. 
    \item Use the MI procedure to create $m$ completed data sets with the missing values replaced by imputed values.
    \item\label{combine} Use each of the $m$ data sets to obtain an estimate of $\theta$ as well as its associated variance and combine them using Rubin's combining rules \citep{rubin2004multiple} to obtain $\hat \theta$ and a 95\% confidence interval (CI). 
    \item Report the bias of $\hat \theta$, the CI interval length and whether or not the CI covered the true value \citep[Section 2.5.2]{vanbuuren:2012}.
\end{enumerate}
We repeat Steps 2-5 $S$ times to obtain the empirical coverage rate. By varying the full data model and the missing data mechanism, in Step 2, we can control the two paths that influence the effectiveness of the MI procedures. 

\subsection{Simulation Study}\label{sec:simulation}
In regression settings, an outcome $Y$ can depend on many explanatory variables $\Z = X_1,\dots, X_J$ some of which can be costly to measure. As such, it is common that while the outcome $Y$ is measured for all variables, some entries of the design matrix $\Z$ are missing. In this simulation, we exclusively focus on this situation and restrict the missingness to the explanatory variables. We will further assume that the missingness mechanism does not allow for the missingness to depend on the outcome $Y$. 

In this situation complete case analysis (or listwise deletion) provides an unbiased estimate of the regression coefficients; however, the reduced sample size often leads to losses in efficiency, through higher standard errors. Another disadvantage of using complete case analysis whenever the number of explanatory variables $J$ is of moderate size is that the probability of having enough complete cases to estimate the regression coefficients is low. In this setting using a MI procedure is paramount and leads to a significant reduction in the standard errors; however, this can induce a slight bias. \cite{white2010bias} show through an extensive simulation study that the increase in bias often time leads to a decreased empirical coverage rate for both \textbf{MAR} and \textbf{MNAR} data sets. 

For our simulation study we set $J=40$, $N=1000$, and consider $X_j$ that include both continuous and discrete variables to demonstrate the versatility of the copula approach without specifying any of the marginal distributions. This is precisely the scenario we described above; the probability of enough complete cases existing to estimate the regression coefficients is effectively 0.\footnote{The reason is that with a high probability of missingness for each variable and a large enough number of variables, the probability of observing all variables for one particular case quickly becomes very small. Specifically, with probability of missingness $p$ and $k$ covariates, the probability of all observations being present for one case is $(1-p)^k$.}
 
The distributions we consider for the elements of the design matrix are Gaussian, Bernoulli, Poisson and ordinal. To make imputation feasible we require the variables to be correlated. To generate correlated variables we first construct a matrix of correlated Gaussian random variables and then transform the variables to have the appropriate marginals. For example, to generate a pair of correlated Poisson random variables $A$ and $B$ with mean $\lambda$ we construct $(Z_1,Z_2)\sim\normal(0,\Sigma)$ where $\sigma_{11}=\sigma_{22}=1$ and $\sigma_{12}=\sigma_{21}=\rho$ and set $A=F^{-1}_{{\rm Pois},\lambda}(F_{\normal}(Z_1))$ and $B=F^{-1}_{{\rm Pois},\lambda}(F_{\normal}(Z_2))$. 
The data generating process thus leads to the following marginal distributions for the entries in $\Z$: for $j=1,\dots,10$
		\begin{align*}
		    X_j&\sim\normal(0,\sigma_j^2)\quad
		    &X_{j+10}&\sim{\rm Bern}(p_j) \\
		    X_{j+20}&\sim{\rm Pois}(\lambda_j)\quad
		    &X_{j+30}&\sim{\rm ordinal}(0,1) \\
		    \Z&=(X_1,\dots,X_{40}) \quad
            &Y&\sim\normal\left(\sum_{i=1}^{40}X_i,1\right),
		\end{align*}
        where  $\sigma_j= 1+(j-1)/9$, $\lambda_j=0.2+2(j-1)/90$ and $p_j= 2+3(j-1)/9$. Both the amount of missingness (MC) and correlation ($\rho$) between the different variables is varied according to the specified values given in Table \ref{T:Sim_Study}.

\begin{table}[t]
\centering
\begin{tabular}{cc}
Correlation ($\rho$) & \shortstack{Missingness\\Coefficient (MC)} \\ \hline\hline
0.2         & 0.3                     \\
0.35        & 0.4                     \\
0.5         & 0.5                     \\
0.65        & 0.6                    
\end{tabular}
\caption{Simulation Study configurations.}
\label{T:Sim_Study}
\end{table}

We consider two missing data mechanisms for $\Z$, one that produces \textbf{MAR} data sets and another one that generates \textbf{MNAR} data sets, see Appendices A and B for details. The MI procedures we considered are only valid under the MAR assumption;  however, it is useful to check how each method performs when this assumption is violated - as is often the case in practice.

\subsection{Results}
We performed 1,000 simulations under each of the possible combinations of the correlation and missingness coefficient, as detailed in Table \ref{T:Sim_Study}, under both \textbf{MAR} and \textbf{MNAR} missing data mechanisms. For \texttt{MICE}, we specified the correct marginal distributions (for example ordered logit model for the ordinal variables). For \texttt{Amelia II}, we used the appropriate variable transformation in accordance with the package help files. For the copula approach, we did not need to specify any distributions/transformations. Using each of the three procedures, we created 20 completed data sets that were used to estimate the regression coefficients and a corresponding 95\% confidence interval.\footnote{Throughout the simulation, the \texttt{Amelia II} software crashed numerous times, as detailed in Table D.1 in Appendix D. Due to this the results for \texttt{Amelia II} are only on a subset of the $36,000$ simulations}. None of the simulations had enough complete cases to estimate the regression coefficients using listwise deletion.

The most significant source of variation in the simulation was due to the different classes of variables, followed by the correlation and the missingness coefficient. There is only a small difference in the results obtained from the \textbf{MAR} and \textbf{MNAR} data; therefore, our discussion will focus on the former, with the figures for the latter included in Appendix C. 
Figures \ref{fig:rho} and \ref{fig:PMis} illustrate how the bias, coverage and interval length, vary across the interaction of the different variable classes, the correlation, and the missingness coefficient, respectively.  Overall the copula method achieved an empirical coverage rate of $93\%$ which was much higher than that of \texttt{MICE}, $87\%$, and \texttt{Amelia II}, $83\%$. Less adversarial regimes were previously studied in \cite{white2010bias}, by reducing the number of covariates in our simulation we can recover similar coverage rates for the MI procedures as are reported there. Both the copula and \texttt{MICE} methods had an absolute average bias of $0.17$. \texttt{Amelia II} performed worse and had a bias of $0.25$. On average, all three methods had approximately the same interval length. 

The copula imputations were obtained using $10,000$ iterations from Hoff's \citeyearpar{hoff2010package} package whose convergence was checked on a subset of simulations. The lag-10 autocorrelation for the thinned chains was less than $0.18$ in absolute value for each of the elements of the latent correlation matrix, and the effective sample size was always above $200$ ($97.6\%$ of the entries were above $500$). Since the copula method is sampling from the posterior distribution which requires the MCMC algorithm to converge to the stationary distribution, its computation time depends on the rate of convergence as well as the desired number of imputations. Running multiple MCMC chains in parallel to generate independent imputations can reduce the computation time. This approach is slightly slower than \texttt{Amelia II} but is substantially faster than the standard \texttt{MICE} algorithm where all $J-1$ variables are used to impute the $j^\text{th}$ variable. Fortunately, the copula algorithm scales well as the sample size and the number of explanatory variables increases. The copula method had the lowest bias, highest coverage rate and often the longest interval length. It is noteworthy that even though the semi-parametric estimation procedure did not require specification of the marginals, any data transformations, or tuning, it still outperformed the other two procedures.

Since the \texttt{MICE} procedure is iterative, users need to check that the model parameters fully explore the parameter space. Unlike the Bayesian copula method, there are no explicit convergence criteria that can be tracked. We performed a visual check that revealed no abnormalities and also ran each \texttt{MICE} chain for $20$ iterations as recommended in \citet{vanbuuren:groothuis-oudshoorn:2011}. 
The \texttt{MICE} method performed almost as well as the copula method but had slightly lower coverage rate, meaning the estimated standard errors were too small. \texttt{MICE} also had the smallest average bias for the normal and Poisson variables. Again, however, these results are contingent on specifying the correct conditional distribution which can often be challenging.

\texttt{Amelia II} had the lowest coverage and highest bias both on average and in most scenarios that we considered. It had the smallest average interval length of $1.23$, which shows that it was systematically underestimating the variance: leading to the low coverage rates. 

Figure \ref{fig:PMis} shows that the average bias and the interval length increases as a function of the proportion of missing values. This leads to a decrease in the empirical coverage as the bias increases at a faster rate than the interval length. One notable exception was the correct coverage of the copula approach for the regression parameters of the ordinal and binomial variables, both \texttt{Amelia II} and \texttt{MICE} undercovered. Given that these types of variables are frequently encountered in social science applications, these results especially suggest that using a copula approach can lead to better statistical conclusions. Moreover, the overall simulation results indicate that when a normal distribution does not well approximate the data, then the copula approach will consistently outperform both \texttt{Amelia II} and \texttt{MICE}.  

Somewhat surprisingly, there seems to be less variation in the bias and the interval length as a function of the correlation, as is shown in Figure \ref{fig:rho}. Except for the normally distributed variables, the bias decreases as the correlation increases due to the reduction in the relative loss of information from the missing data. 

\begin{figure}[htbp]
	\centering
	\includegraphics[scale=.6]{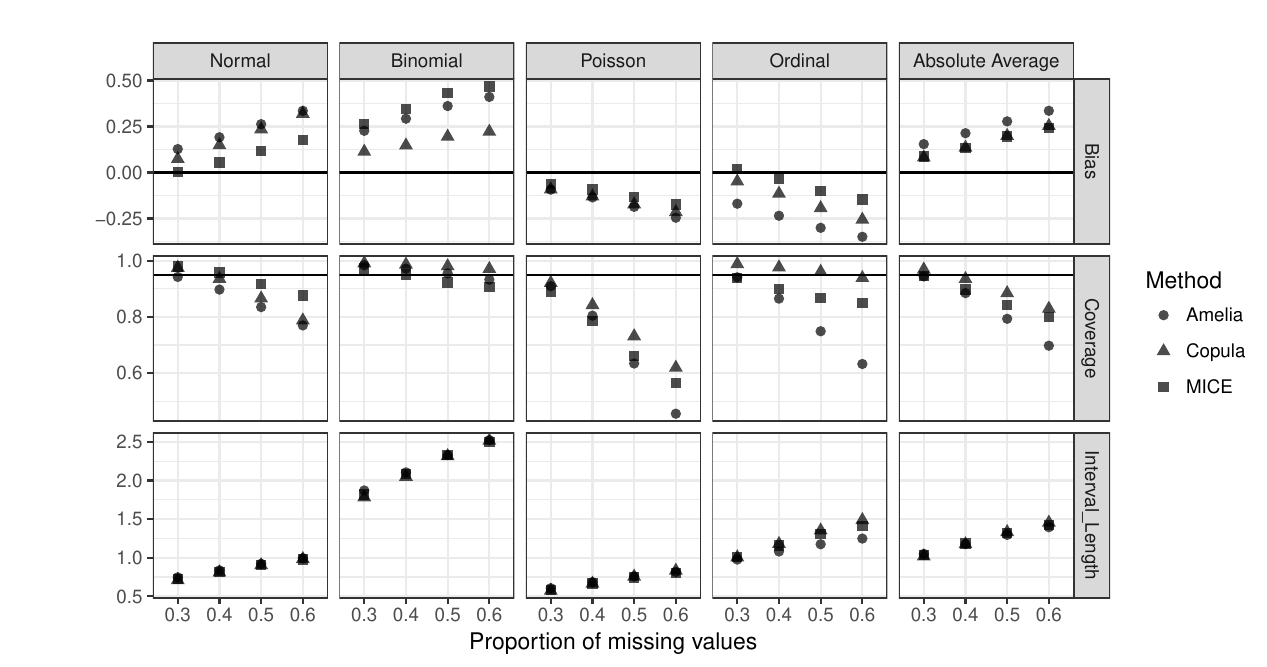}
	\caption{Simulation study results for the \textbf{MAR} data as a function of the missingness coefficient, averaging over the correlation. The plot is split by the different variable types (normal, binomial, Poisson and ordinal) and the three outcomes of interest (bias, coverage and interval length). The rightmost panel shows the result averaging over the different variable types.}
	\label{fig:PMis}
\end{figure}

\begin{figure}[htbp]
	\centering
	\includegraphics[scale=.6]{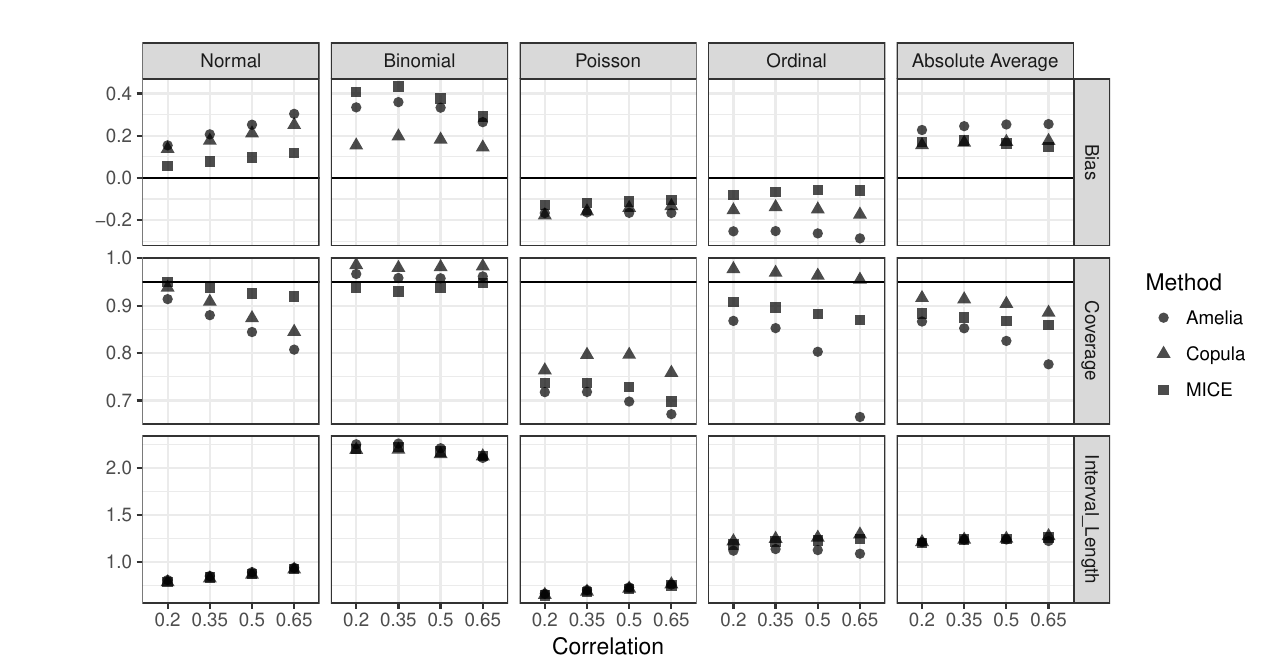}
	\caption{Simulation study results for the \textbf{MAR} data as a function of the correlation, averaging over the missingness coefficient. The plot is split by the different variable types (normal, binomial, Poisson and ordinal) and the three outcomes of interest (bias, coverage and interval length). The rightmost panel shows the result averaging over the different variable types.}

	\label{fig:rho}
\end{figure}

Breaking the \textbf{MAR} assumption did not lead to drastically worse results. We observe a decrease of about $3\%$ in the coverage of all three methods and a slight decrease in the average bias. This shows that the methods are somewhat robust to violations of MAR assumption when it is not too severe. Figures C.1 and C.2 in the Appendix C show the results of the simulations when the MAR assumption is violated.

\clearpage

\section{Application Study}\label{sec:application}

In this section, we provide a comparison of the three imputation methods using an application from political science. The empirical example shows how copula methods can be used to generate imputations in a large data set with a variety of variable types. 

\subsection{Inequality and Democratic Support}
As we have elaborated above, imputation methods are still underused, especially in the social sciences. There is, however, some visible progress. One example where scholars have taken advantage of one of the imputation methods currently available is ``Economic Inequality and Democratic Support'' by \citet{krieckhaus:etal:2014} published in the \textit{Journal of Politics}. \citet{krieckhaus:etal:2014} explore whether the support for democracy within countries is affected by the level of inequality. The authors combine country level variables (such as inequality) with individual level survey data from $40$ democracies around the world. For multiple countries several survey waves are included, resulting in $57$ country-years and a total of $77,642$ observations \citep[144]{krieckhaus:etal:2014}. For this replication exercise we replicate \textit{Model $1$} in \textit{Table 1} in \citet{krieckhaus:etal:2014}. The dependent variable is a ``$13$-point additive index (ranging from $0$ to $12$) of democratic support'', which the authors treat as a continuous variable \citep[144]{krieckhaus:etal:2014}. The main independent variables of interest are \textit{Inequality} at the country level, and an ordinal \textit{Income} scale at the individual level (ranging from $1$ to $10$). Additionally, the authors control for \textit{Age}, \textit{Gender}, \textit{Institutional Confidence}, \textit{Interest in Politics}, \textit{Interpersonal Trust}, \textit{Education}, \textit{Prior Regime Evaluation}, and \textit{Leftist Ideology} all drawn from the \textit{World Values Survey} \citep{WVS:2012}. As in the original article, all individual level variables are demeaned ``using group-mean centering'' after the imputation \citep[145]{krieckhaus:etal:2014}. The data are analyzed using a random-coefficients model.

\begin{table}[!htbp] \centering 
  \caption{Share of Missingness in Variables of Interest} 
  \label{appl:missing} 
\scalebox{0.8}{
\begin{tabular}{@{\extracolsep{5pt}} cccc} 
\\[-1.8ex]\hline 
\hline \\[-1.8ex] 
Democracy Support & Inequality & Income & Age \\ 
19.9 & 1.8 & 12.9 & 0.2 \\ 
[1.8ex] 
Gender & Institutional
 Confidence & Interest in
 Politics & Interpersonal
 Trust \\ 
0.1 & 11.7 & 2.5 & 3.7 \\
[1.8ex] 
Education & Leftist
 Ideology & Prior Regime
 Evaluation &  \\ 
3.9 & 18.5 & 21.3 &  \\ 
\hline \\[-1.8ex] 
\end{tabular} 
}
\end{table} 

Most importantly for this study, the original data suffers from a relatively high number of missing observations. Table \ref{appl:missing} shows the share of missing observations for variables included in the replication exercise. We can see that many of the variables have a large share of missing observations. If instead of multiple imputations, the authors used in listwise deletion then the number of observation in the regression model would have been approximately halved. Instead, \citet{krieckhaus:etal:2014} use \texttt{Amelia II} to multiple impute five data sets which they analyze. Estimates are then combined using Rubin's rule. 

This is an excellent setting for our comparison of multiple imputation techniques. The number of missing observations is quite large, and the data set includes different types of variables, continuous, binary, as well as ordinal. We created $20$ multiple imputed data sets using each of the imputation techniques: \texttt{Amelia II}, \texttt{MICE}, and \texttt{sbgcop}. We then re-estimate \textit{Model 1} in \textit{Table 1} in \citet[147]{krieckhaus:etal:2014} and combine the estimation results for each method's multiple imputed data sets via Rubin's rule. 

For \texttt{Amelia II} we specify the type of each variable and then generate 20 imputed data sets using the full original data. Similarly, we declare each variable's type for \texttt{MICE} and estimate the default model for each. We use all variables except the one to be imputed as independent variables in the chained equations. Again, we create $20$ multiple imputed data sets and set the maximum number of iterations to $20$. 

Lastly, we use our preferred method, imputation via the semi-parametric Gaussian copula, to generate $20$ imputed data sets. We run the MCMC chain for $2100$ iterations and randomly draw 20 data sets from the posterior. Note that, again, we do not have to declare any of the variable types or make any other specification or transformation of the data. 

\begin{figure}[htbp]
\centering
\includegraphics[scale=.5]{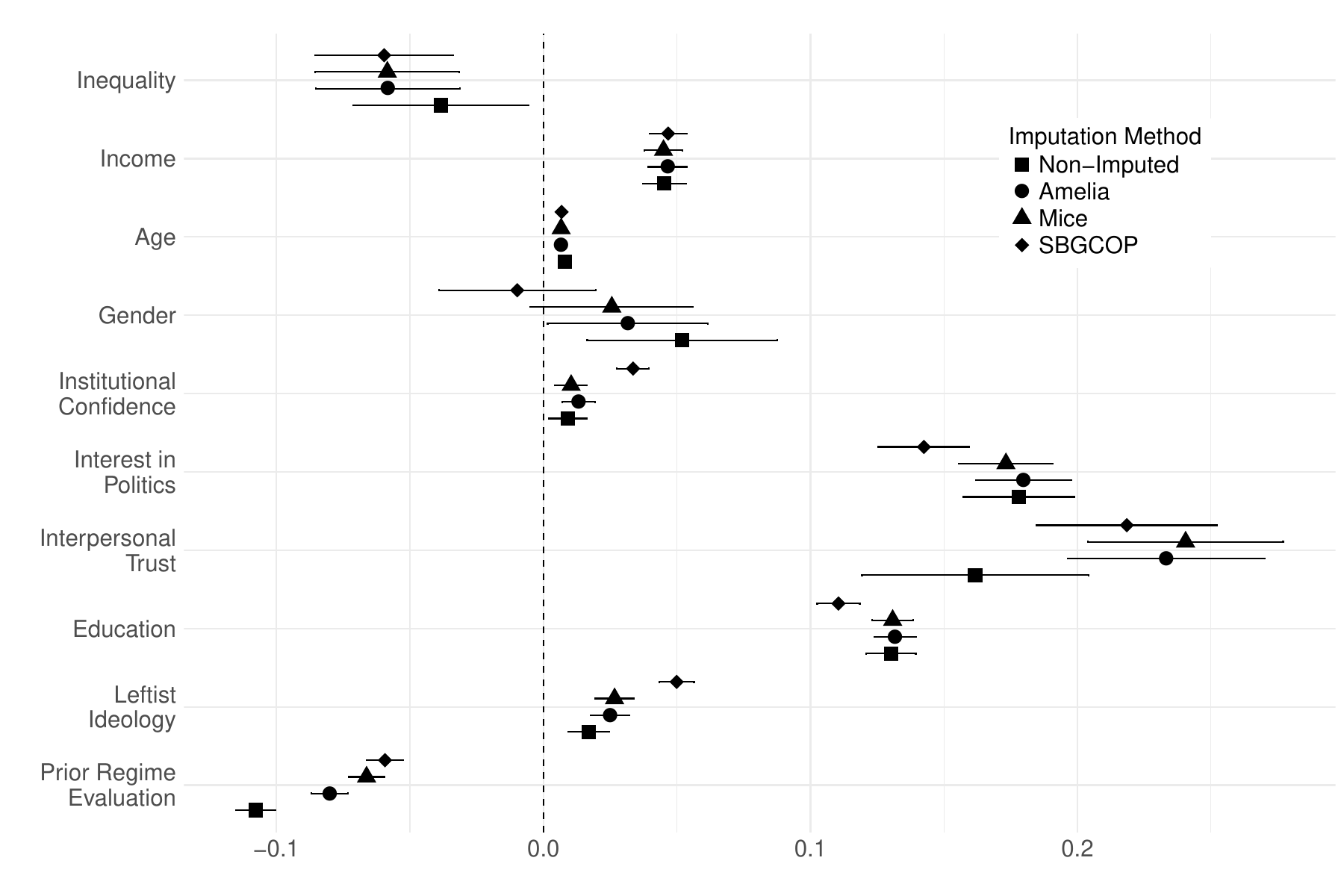}
\caption{Coefficient estimates and confidence intervals for \textit{Model 1} in \textit{Table 1} in \citet{krieckhaus:etal:2014} based on three imputation techniques and list-wise deletion}
\label{app:results}
\end{figure}

Figure \ref{app:results} shows the coefficient estimates and $95\%$ intervals for the replicated model based on each of the imputation techniques, as well as when list-wise deletion is used. First, the results are quite similar for the \emph{Inequality}, \emph{Income}, and \emph{Age} variables. Even for the models based on listwise deletion. For the two main variables of interest, inequality, and income, the results based on different imputation techniques are virtually the same. 

On the other hand, there are several significant differences for the other variables included in the model. First, the effect of gender is essentially zero according to the models estimated on the copula imputed data. Based on the data imputed using \texttt{MICE} or \texttt{Amelia II}, females have higher ratings of democracy satisfaction (though the confidence intervals just cover zero). According to the non-imputed data, the effect of gender is quite strong.

Based on the data imputed with the copula method, the estimated association of \textit{Institutional Confidence} with \textit{Democracy Satisfaction} is significantly stronger compared to the models based on listwise deletion or other imputation methods. Similarly, the estimated effect of \textit{Leftist Ideology} is also substantially larger according to the copula imputed data. On the other hand, the association of \textit{Education} levels with \textit{Democracy Satisfaction} is significantly smaller. Based on the copula, the relationships of \textit{Interest in Politics}, and \textit{Prior Regime Evaluation} with the dependent variable of \textit{Democracy Satisfaction} are all modeled to be weaker, compared to the other methods (and the non-imputed data), though the confidence intervals overlap.

It is interesting to note, that, except for one variable (\textit{Interpersonal Trust}), whenever the estimated coefficient for the copula imputed data differs from the coefficients based on the other imputation methods, it is in the opposite direction of the difference to the list-wise deletion coefficient. This is especially easy to see for the \textit{Gender} and \textit{Leftist Ideology} variables, where the effect is strongest (weakest) according to the model estimated on the list-wise deleted data and weakest (strongest) for the copula based models.

Based on the simulation results, especially with respect to binary and ordinal variables, and the theoretical properties we are confident in the accuracy of the copula imputation method. These results suggest then that \textit{Gender} is not associated with people's satisfaction with democracy, whereas \textit{Institutional Confidence} and \textit{Left} ideology both have much stronger effects. 

\clearpage
\section{Conclusion}

What practical lessons can we learn about how to deal with missing data?  
In this article, we re-emphasize the importance of dealing with missing data and present a copula based approach, developed by \citet{hoff:2007}, that is elegant and requires little pre-specification of the data. With the rank based approach introduced by \citet{hoff:2007}, the Gaussian copula can be used to impute binary, ordinal, and continuous variables. We discuss the theoretical properties of the copula method and its theoretical attractiveness compared to other commonly employed techniques. In particular, the Gaussian copula introduced here enables researchers to make imputation via draws from a valid posterior of the joint distribution without specifying the distributions of the individual variables. Moreover, we present evidence from simulations that it performs better than either \texttt{Amelia II} or \texttt{MICE}, especially when it comes to non-normally distributed data. 

While the three imputation methods perform relatively similarly, throughout the simulation, the Copula method does have the lowest average bias (tied with \texttt{MICE}) and the highest coverage rate ($93\%$). More so, \texttt{MICE} requires specification of the conditional distributions whereas the copula method does not. Recent theoretical results for \texttt{MICE} suggest that good performance heavily relies on being approximately correct in the choice of conditionals \citep{li2012imputing}. On the other hand, theoretical guarantees for good behavior of copula methods are available. In particular, information bounds for rank-based estimators are the same as the information bounds for estimators based on the full (scale and rank) data \citep{hoff2014information}. Under \textbf{MAR} and \textbf{MCAR} we inherit all the properties of the full data, and by introducing structure to the imputation, we are likely to have good behavior even under 
\textbf{MNAR}.

One aspect that we have not addressed herein is the validity and sensitivity to the unassessable assumptions made when analyzing data with missing values \citep{molenberghs2014handbook}, \emph{i.e.} the type of missingness mechanism. \citet{bojinov2017diagnosing} show that the Gaussian copula approach can be used to assess the validity of the missing always at random assumption (a slightly stronger assumption that implies \textbf{MAR}). Their results suggest that by using a Gaussian copula for generating imputations, the analyst can also easily diagnose the assumptions they made and quickly identify variables which are likely to break these assumptions. 
This adds another benefit to using the method discussed in this paper. 

Consideration must also be given to the computational cost of any procedure.
As indicated by \citet{graham:2009} the disadvantages of EM approaches are especially large when imputing databases with many variables or applications of ``big data''. \texttt{MICE}  can be computationally less expensive but suffers when the number of variables increases as the correct choice for each of the conditionals becomes increasingly unlikely. 
The semiparametric copula approach described here relies on MCMC, its speed does not depend on the fraction of missing data and scales nicely in the dimension of the dataset. This makes it possible to impute even large database in a relatively timely manner and no pre-specification of the data. Moreover, using the copula model to multiply impute missing values provides some of the advantages (such as a proper posterior distribution of the data) but is less burdensome on scholars than imputing values in a fully Bayesian approach \citep{erler2016dealing}.

Finally, the copula approach is quite flexible and can be employed at different stages of the analysis process. First, it can be used to generate a single estimate of the missing data or the mean of a large number of draws, which is exactly what might be needed in some situations. Second, per the recommendation of Rubin, it can be used to construct multiple databases. As with \texttt{Amelia II}, the copula imputations can be analyzed separately and the results combined using either \texttt{mitools} or \texttt{Zelig} \citep{imai:etal:2008} in $\mathcal{R}$. 
Thus, the copula approach to missing data can be explicitly integrated into the modeling and analysis of observational data in a simplistic, organic fashion.


\clearpage
\bigskip
\begin{center}
{\large\bf SUPPLEMENTARY MATERIAL}
\end{center}
Code will be provided on the author's Dataverse. 
\begin{description}
\item[R-packages for Imputation:] 3 R-packages used to impute the missing data: Amelia II, MICE, sbgcop 
\item[R-code for simulation in Section 4:] R-code to replicate simulation study in section 4.
\item[R-code for Application in Section 5:] R-code to replicate application in section 5. 
\end{description}

\clearpage

\spacingset{2} 

\appendix\label{app:mechanism}
\setcounter{table}{0}
\renewcommand{\thetable}{\Alph{section}.\arabic{table}}

\setcounter{figure}{0}
\renewcommand{\thefigure}{\Alph{section}.\arabic{figure}}

\section{Missing at Random}

We now describe a missing data mechanism that always produces \textbf{MAR} data. Our goal is to make the simulations as realistic as possible; therefore some variables will be fully observed, and others will have different amounts of missing values. 

\begin{enumerate}
	\item Given a fully observed data set $\Z$ randomly select four variables, one from each of the four classes, that will be fully observed; without loss of generality relabel them $X_1,X_{11},X_{21}$ and $X_{31}$.  
    \item Randomly select four variables from the remaining thirty six, one from each of the four classes, that will have a 5-6\% missingness; without loss of generality relabel them $X_2,X_{12},X_{22}$ and $X_{32}$. The probability that the $i^\text{th}$ observation for each variable is missing is based on a logistic regression on the fully observed variables, $X_1,X_{11},X_{21}$ and $X_{31}$, adjusted so that the mean number of missing variables is between 5-6\%. The missingness indicators are then sampled from independent Bernoulli random variables with the appropriate probabilities.  Let $\Z^{(1)}=(X_1,X_2,X_{11},X_{12},X_{21},X_{22},X_{31},X_{32})$ and $\Z^{(1)}_\text{cc}$ be the complete cases after removing the any rows that have missing values. 
    \item The probability of the $i^\text{th}$ observation missing for the remaining thirty two variables is proportional to a logistic regression on the fully observed $\Z^{(1)}_\text{cc}$. The probabilities are then adjusted so that the mean number of missing variables is equal to the Missingness Coefficient (MC) (see Table 1 for the range of values that we considered). The missingness indicators are sampled from independent Bernoulli random variables with the appropriate probabilities. If the $i^\text{th}$ row of $\Z^{(1)}$ has been removed in $\Z^{(1)}_{cc}$ then that row is always observed for the thirty-two variables. 
\end{enumerate}
The proportion of missing values is slightly lower than the MC as four variables are fully observed, and four others only have 5-6\% of their values missing. 
\section{Missing not at Random}
		
We now describe a missing data mechanism that produces \textbf{MNAR} data with extremely high probability. 
        
 \begin{enumerate}
	\item Given a fully observed data set $\Z$ randomly select four variables, one from each of the four classes, that will be fully observed; without loss of generality relabel them $X_1,X_{11},X_{21}$ and $X_{31}$.  
    \item Randomly select four variables from the remaining thirty six, one from each of the four classes, that will have a small amount of missingness; without loss of generality relabel them $X_2,X_{12},X_{22}$ and $X_{32}$. The probability that the $i^\text{th}$ observation is missing is given by, 
    \begin{align*}
		P(R_2=1|\Z) &= 1_{X_2>0} p_{MC}, \\
        P(R_{12}=1|\Z) &= 1_{X_{12}=0}  p_{MC}, \\
		P(R_{22}=1|\Z) &= 1_{X_{22}>3}  p_{MC}, \\
        P(R_{32}=1|\Z) &= 1_{X_{32}=3}  p_{MC}, \\        
	\end{align*}
    where the value of $p_{MC}$ is given by the MC in Table 1.
    \item For the remaining thirty two variables the probability of the $i^\text{th}$ observation missing is based on a logistic regression on $\Z^{(1)}$ adjusted so that the mean number of missing variables is equal to the MC (see Table 1). The missingness indicators are again sampled from independent Bernoulli random variables with the appropriate probabilities. In contrast to the MAAR mechanism if the $i^\text{th}$ row of $\Z^{(1)}$ has missing values then other variables in that row can still be missing.  
\end{enumerate}

\clearpage 

\section{Plots of \textbf{MNAR} Simulation Results}
\begin{figure}[htbp]
	\centering
   	\includegraphics[scale=.6]{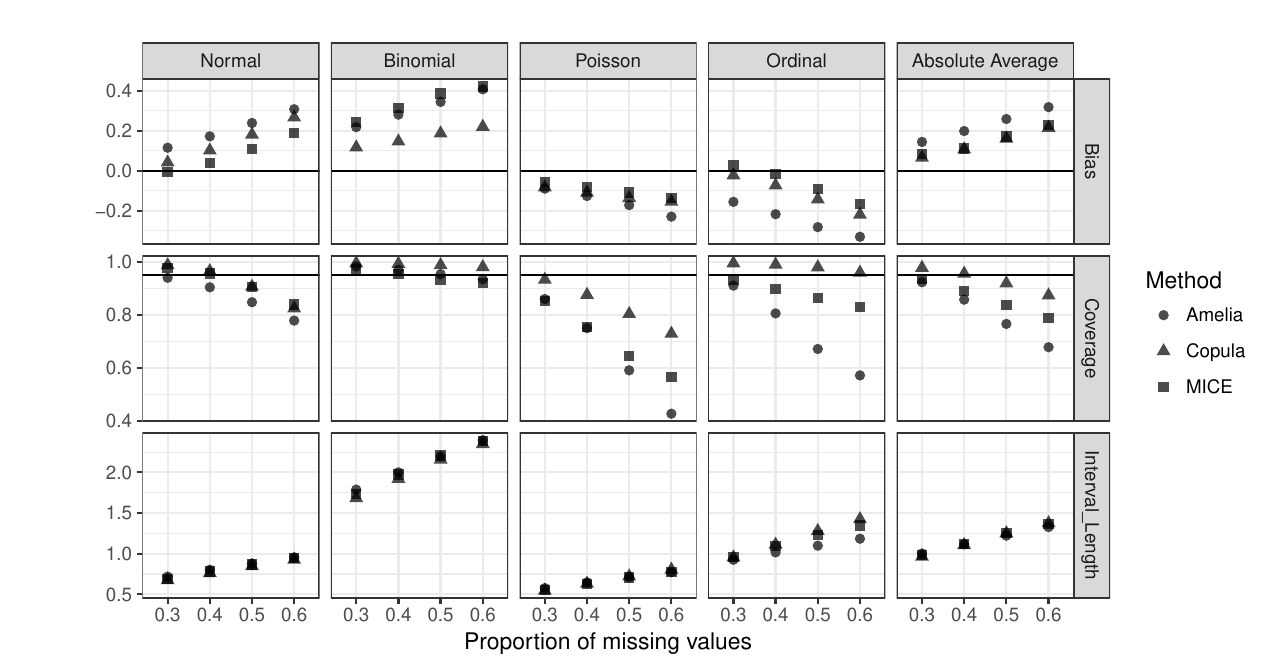}
	\caption{Simulation study results for the \textbf{MNAR} data as a function of the missingness coefficient, averaging over the correlation. The plot is split by the different variable types (normal, binomial, Poisson and ordinal) and the three outcomes of interested (the bias, coverage and interval length). The rightmost panel shows the result averaging over the different variable types.}
	\label{fig:PMis_mnar}
\end{figure}

\begin{figure}[htbp]
	\centering
   	\includegraphics[scale=.6]{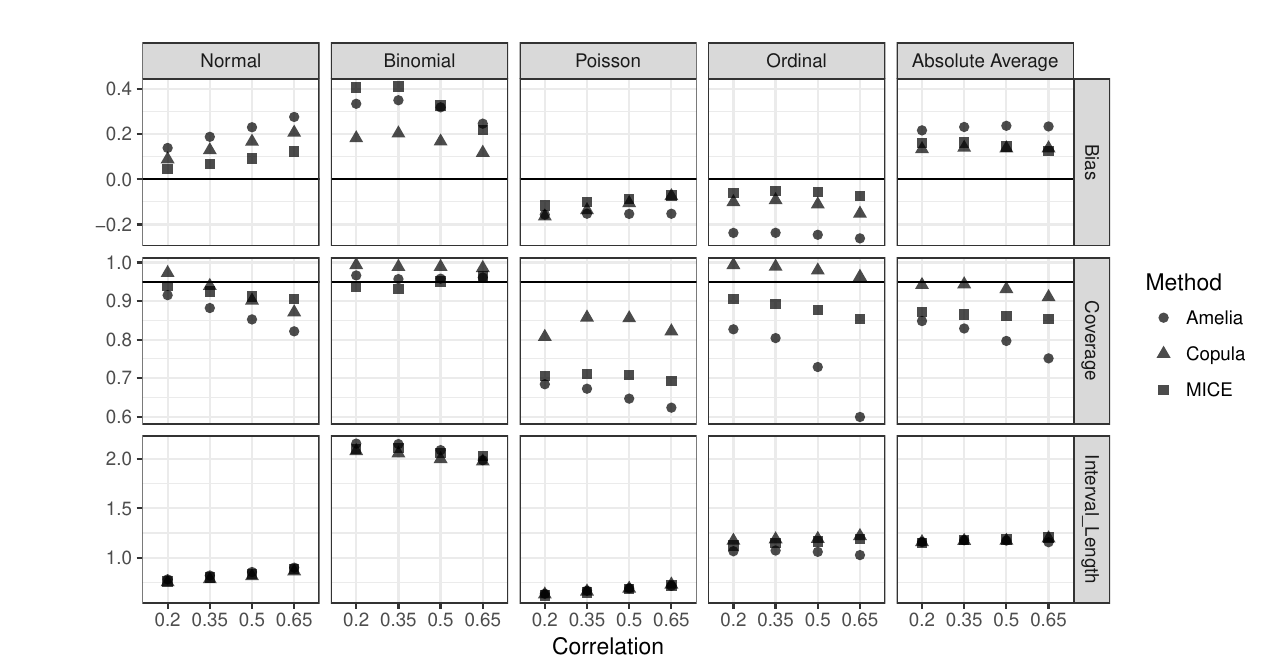}
	\caption{Simulation study results for the \textbf{MNAR} data as a function of the correlation, averaging over the missingness coefficient. The plot is split by the different variable types (normal, binomial, Poisson and ordinal) and the three outcomes of interested (the bias, coverage and interval length). The rightmost panel shows the result averaging over the different variable types.}

	\label{fig:rho_mnar}
\end{figure}

\clearpage

\section{Number of Simulations for which \texttt{Amelia II} crashed}

\begin{table}[h!]
\centering
\label{tab:Kill_Amelia}
\begin{tabular}{cccccc}
\cmidrule[\heavyrulewidth]{3-6}
& & \multicolumn{4}{c}{Correlation }                                               \\
&                          & 0.2 & 0.35 & 0.5 & 0.65 \\ 
\cmidrule(r){3-6}
\cmidrule(r){3-6}
                      & \multicolumn{1}{c}{0.3} & 2   & 0    & 0   & 7    \\
          Share of            & \multicolumn{1}{c}{0.4} & 93  & 16   & 8   & 0    \\
            Missingness          & \multicolumn{1}{c}{0.5} & 285 & 138  & 37  & 13   \\
                      & \multicolumn{1}{c}{0.6} & 485 & 305  & 159 & 72  
\end{tabular}
\caption{The number of \texttt{Amelia II} crashes out of the 1000 simulations under each of the possible scenarios.}
\end{table}

\clearpage
\bibliographystyle{chicago2}
\bibliography{master}
\end{document}